\documentclass[preprint,12pt]{articletonppj}
\usepackage{graphicx}

\usepackage{amssymb}

\newcommand\rf[1]{(\ref{eq:#1})}
\newcommand\lab[1]{\label{eq:#1}}
\newcommand\nonu{\nonumber}
\newcommand\br{\begin{eqnarray}}
\newcommand\er{\end{eqnarray}}
\newcommand\be{\begin{equation}}
\newcommand\ee{\end{equation}}

\newcommand\lb{\lbrack}
\newcommand\rb{\rbrack}
\newcommand\llangle{\left\langle}
\newcommand\rrangle{\right\rangle}

\newcommand\llb{\left\lbrack}
\newcommand\rrb{\right\rbrack}

\renewcommand\({\left(}
\renewcommand\){\right)}
\newcommand\bv{\bigm\vert}               

\newcommand\bc{\begin{center}}
\newcommand\ec{\end{center}}

\newcommand\partder[2]{\frac{{\partial {#1}}}{{\partial {#2}}}}

\renewcommand\d{\delta}

\newcommand\eps{\epsilon}
\newcommand\vareps{\varepsilon}
\newcommand\g{\gamma}
\newcommand\G{\Gamma}

\newcommand\h{\frac{1}{2}}
\renewcommand\k{\kappa}
\renewcommand\l{\lambda}
\renewcommand\L{\Lambda}
\newcommand\m{\mu}
\newcommand\n{\nu}

\newcommand\vp{\varphi}

\newcommand\pa{\partial}

\newcommand\pr{\prime}

\newcommand\s{\sigma}

\renewcommand\t{\tau}
\renewcommand\th{\theta}

\newcommand\cA{{\mathcal A}}

\newcommand\cF{{\mathcal F}}

\newcommand\cM{{\mathcal M}}

\newcommand{\ct}[1]{\cite{#1}}
\newcommand{\bib}[1]{\bibitem{#1}}
\newcommand\PRL[3]{\textsl{Phys. Rev. Lett.} #2; #1: #3}
\newcommand\NPB[3]{\textsl{Nucl. Phys.} #2; B#1; #3}

\newcommand\PRD[3]{\textsl{Phys. Rev.} #2; D#1; #3}

\newcommand\PLB[3]{\textsl{Phys. Lett.} #2; #1B: #3}
\newcommand\CQG[3]{\textsl{Class. Quantum Grav.} #2; #1: #3}

\newcommand\AoP[3]{\textsl{Ann. of Phys.} #2; #1: #3}

\newcommand\IJMPA[3]{\textsl{Int. J. Mod. Phys.} #2; A#1: #3}

\newcommand\MPLA[3]{\textsl{Mod. Phys. Lett.} #2; A#1: #3}

\begin{document}

\begin{frontmatter}



\title{Hiding Charge in a Wormhole}


\author[BGU]{Eduardo Guendelman\corref{cor1}}
\ead{guendel@bgu.ac.il}
\cortext[cor1]{Corresponding author -- tel. +972-8-647-2508, fax +972-8-647-2904.}
\author[BGU]{Alexander Kaganovich}
\ead{alexk@bgu.ac.il}
\address[BGU]{Department of Physics, Ben-Gurion University of the Negev,
P.O.Box 653, IL-84105 ~Beer-Sheva, Israel}

\author[INRNE]{Emil Nissimov}
\ead{nissimov@inrne.bas.bg}
\author[INRNE]{Svetlana Pacheva}
\ead{svetlana@inrne.bas.bg}
\address[INRNE]{Institute for Nuclear Research and Nuclear Energy, Bulgarian Academy
of Sciences, Boul. Tsarigradsko Chausee 72, BG-1784 ~Sofia, Bulgaria}

\begin{abstract}
Existence of wormholes can lead to a host of new effects like Misner-Wheeler
``charge without charge'' effect, where without being generated 
by any source an electric flux arrives from one ``universe'' and flows into the other
``universe''. Here we show the existence of an intriguing opposite possibility. 
Namely, a charged object (a charged lightlike brane in our case) sitting at the wormhole 
``throat'' expels all the flux it produces into just one of the ``universes'', 
which turns out to be of compactified (``tube-like'') nature. An outside
observer in the non-compact ``universe'' detects, therefore, a neutral object.
This charge-hiding effect takes place in a gravity/gauge-field system
self-consistently interacting with a charged lightlike brane as a matter
source, where the gauge field subsystem is of a special non-linear form
containing a square-root of the Maxwell term and which previously has been
shown to produce a QCD-like confining gauge field dynamics in flat space-time. 
\end{abstract}


\end{frontmatter}


\section{Introduction}
\label{intro}

Misner-Wheeler ``charge without charge'' effect \ct{misner-wheeler} stands out 
as one of the most interesting physical phenomena produced by wormholes.
Misner and Wheeler realized that wormholes connecting two asymptotically flat 
space times provide the possibility of existence of electromagnetically
non-trivial solutions, where the lines of force of the electric field flow from one 
universe to the other without a source and giving the impression of being 
positively charged in one universe and negatively charged in the other universe.
For a detailed account of the general theory of wormholes we refer to Visser's book 
\ct{visser-book} (see also \ct{visser-1,visser-2} and some more recent accounts 
\ct{WH-rev-1}--\ct{bronnikov-2}.

In the present paper we find the opposite effect in wormhole physics,
namely, that a genuinely charged matter source of gravity and electromagnetism
may appear {\em electrically neutral} to an external observer. Here we show this 
phenomenon to take place in a gravity/gauge-field system
self-consistently coupled to a charged lightlike brane as a matter
source, where the gauge field subsystem is of a special non-linear form
containing a square-root of the Maxwell term. The latter has been previously shown 
\ct{GG-1}--\ct{GG-6} to produce a QCD-like confining (``Cornell'' 
\ct{cornell-potential-1}--\ct{cornell-potential-3}) potential in flat space-time. 
In the present case the lightlike brane, which connects as a wormhole ``throat'' 
a non-compact ``universe'' with a compactified ``universe'', is electrically charged, 
however all of its flux flows into the compactified (``tube-like'') ``universe'' only. 
No Coulomb field is produced in the non-compact ``universe'', therefore, the 
wormhole hides the charge from an external observer in the latter ``universe''.

Let us recall that lightlike branes are singular null (lightlike) 
hypersurfaces in Riemannian space-time which provide dynamical description of 
various physically important  phenomena in cosmology and astrophysics such as:
(i) impulsive lightlike signals arising in cataclysmic astrophysical events 
(supernovae, neutron star collisions) \ct{barrabes-hogan}; 
(ii) dynamics of horizons in black hole physics -- the so called ``membrane paradigm''
\ct{membrane-paradigm};
(iii) the thin-wall approach to domain walls coupled to 
gravity \ct{Israel-66}--\ct{Berezin-etal}. 

The gravity/gauge-field system with a square-root of the Maxwell term 
was recently studied in \ct{grav-cornell} (see the brief review in Section 2
below) where the following interesting new
features of the pertinent static spherically symmetric solutions have been found:

(i) appearance of a constant radial electric field (in addition to the Coulomb one)
in charged black holes within Reissner-Nordstr{\"o}m-de-Sitter-type 
and/or Reissner-Nordstr{\"o}m-{\em anti}-de-Sitter-type 
space-times, in particular, in electrically neutral black holes with 
Schwarzschild-de-Sitter 
and/or Schwarzschild-{\em anti}-de-Sitter 
geometry;

(ii) novel mechanism of {\em dynamical generation} of cosmological constant
through the nonlinear gauge field dynamics of the ``square-root'' Maxwell term;

(iii) appearance of confining-type effective potential in charged test particle 
dynamics in the above black hole backgrounds.

In Section 3 of the present paper we extend the analysis in \ct{grav-cornell}
by finding new solutions of Levi-Civita-Bertotti-Robinson type
\ct{LC-BR-1}--\ct{LC-BR-3}, \textsl{i.e.}, 
with space-time geometry of the form $\cM_2 \times S^2$ with $\cM_2$ being a
two-dimensional anti-de Sitter, Rindler or de Sitter space depending on the
relative strength of the electric field w.r.t. coupling of the square-root 
Maxwell term.

In our previous papers \ct{LL-main-1}--\ct{beograd-2010} we have 
provided an explicit reparametrization invariant world-volume Lagrangian formulation of
lightlike $p$-branes (a brief review is given in Section 4)
and we have used them to construct various types of wormhole, 
regular black hole and lightlike braneworld solutions in $D\!=\!4$ or 
higher-dimensional asymptotically flat or asymptotically anti-de Sitter bulk 
space-times. In particular, in refs.\ct{BR-kink}--\ct{beograd-2010} we have
shown that lightlike branes can trigger a series of spontaneous
compactification-decompactification transitions of space-time regions,
\textsl{e.g.}, from ordinary compactified (``tube-like'') 
Levi-Civita-Bertotti-Robinson space to non-compact Reissner-Nordstr{\"o}m or 
Reissner-Nordstr{\"o}m-de-Sitter region or {\sl vice versa}. 
Let us note that wormholes with ``tube-like'' structure (and regular black holes with 
``tube-like'' core) have been previously obtained within different contexts in 
refs.\ct{eduardo-wh}--\ct{zaslavskii-3}.

The essential role of the above mentioned proper world-volume Lagrangian formulation of
lightlike branes manifests itself most clearly in the correct self-consistent 
construction \ct{LL-main-5,Kerr-rot-WH-2} of the simplest wormhole solution first 
proposed by Einstein and Rosen \ct{einstein-rosen} -- the Einstein-Rosen ``bridge'' wormhole.
Namely, in refs.\ct{LL-main-5,Kerr-rot-WH-2} it
has been shown that the Einstein-Rosen ``bridge'' in its original formulation
\ct{einstein-rosen} naturally arises as the simplest particular case of {\em static} 
spherically symmetric wormhole solutions produced by lightlike branes as
gravitational sources, where the two identical ``universes'' with Schwarzschild
outer-region geometry are self-consistently glued together by a lightlike brane occupying
their common horizon -- the wormhole ``throat''. An understanding of this
picture within the framework of Kruskal-Szekeres manifold was subsequently
provided in ref.\ct{poplawski}, which involves Rindler's elliptic
identification of the two antipodal future event horizons.

At this point let us strongly emphasize that the original notion of 
``Einstein-Rosen bridge'' in ref.\ct{einstein-rosen} is qualitatively different from 
the notion of ``Einstein-Rosen bridge'' defined in several popular textbooks ({\sl e.g.}, 
refs.\ct{MTW,Carroll}) using the Kruskal-Szekeres manifold, where the ``bridge'' has 
{\em dynamic} space-time geometry. Namely, the two regions in 
Kruskal-Szekeres space-time corresponding to the two copies of outer Schwarzschild 
space-time region ($r>2m$) (the building blocks of the original static Einstein-Rosen 
``bridge'') and labeled $(I)$ and $(III)$ in ref.\ct{MTW} are generally
{\em disconnected} and share only a two-sphere (the angular part) as a common border
($U=0, V=0$ in Kruskal-Szekeres coordinates), whereas in the original Einstein-Rosen
``bridge'' construction \ct{einstein-rosen} the boundary between the two identical 
copies of the outer Schwarzschild space-time region ($r>2m$) is a three-dimensional 
lightlike hypersurface ($r=2m)$. 

In Section 5 below we consider self-consistent coupling of gravity/gauge-field 
system with a square-root of the Maxwell term to a charged lightlike brane, 
which will serve as a matter source of gravity and (nonlinear) electromagnetism. 
In this Section we derive the main result of the present paper -- wormhole-like 
solutions joining a non-compact ``universe'' to a compactified (``tube-like'')
``universe'' (of generalized Levi-Civita-Bertotti-Robinson type) via a wormhole 
``throat'' realized by the charged lightlike brane, which completely hides
its electric flux from an outside observer in the non-compact ``universe''.
This new charge ``confining'' phenomena is entirely due to the presence of
the ``square-root'' Maxwell term.

\section{Lagrangian Formulation. Spherically Symmetric Solutions}
\label{lagrange}
We will consider the simplest coupling to gravity of the nonlinear gauge field system
with a square-root of the Maxwell term known to produce QCD-like confinement
in flat space-time \ct{GG-1}--\ct{GG-6}. The relevant action is given by (we use units with 
Newton constant $G_N=1$):
\br
S = \int d^4 x \sqrt{-G} \Bigl\lb \frac{R(G)}{16\pi} + L(F^2)\Bigr\rb \quad ,\quad
L(F^2) = - \frac{1}{4} F^2 - \frac{f}{2} \sqrt{\vareps F^2} \; ,
\lab{gravity+GG} \\
F^2 \equiv F_{\k\l} F_{\m\n} G^{\k\m} G^{\l\n} \quad ,\quad 
F_{\m\n} = \pa_\m A_\n - \pa_\n A_\m \; .
\nonu
\er
Here $R(G)$ is the scalar curvature of the space-time metric
$G_{\m\n}$ and $G \equiv \det\Vert G_{\m\n}\Vert$; the sign factor $\vareps = \pm 1$
in the square root term in \rf{gravity+GG} corresponds to ``magnetic'' or ``electric''
dominance; $f$ is a positive coupling constant. It is important to stress that 
we will {\em not} introduce any bare cosmological constant term.

Let us note that the Lagrangian $L(F^2)$ in \rf{gravity+GG} contains both the 
usual Maxwell term as well as a non-analytic function of $F^2$ and thus it
is a {\em non-standard} form of nonlinear electrodynamics. In this way it is 
significantly different from the original purely ``square root'' Lagrangian 
$- \frac{f}{2}\sqrt{F^2}$ first proposed by Nielsen and Olesen \ct{N-O-1} to
describe string dynamics (see also refs.\ct{N-O-2,N-O-3}).
The natural appearance of the ``square-root'' Maxwell term in effective
gauge field actions was further motivated by `t Hooft \ct{tHooft-03}
who has proposed that such gauge field actions are adequate for describing 
confinement (see especially Eq.(5.10) in \ct{tHooft-03}). He has in particular described a
consistent quantum approach in which ``square-root'' gauge-field terms play the
role of ``infrared counterterms''.
Also, it has been shown in first three refs.\ct{GG-1}--\ct{GG-6} that the square root of 
the Maxwell term naturally arises as a result of spontaneous breakdown of scale symmetry of 
the original scale-invariant Maxwell theory with $f$ appearing as an integration 
constant responsible for the latter spontaneous breakdown.

Let us also remark that one could start with the non-Abelian version of the 
gauge field action in \rf{gravity+GG}. Since we will be interested in static 
spherically symmetric solutions, the non-Abelian gauge theory effectively reduces 
to an Abelian one as pointed out in the ref.\ct{GG-1}.

The corresponding equations of motion read:
\be
R_{\m\n} - \h G_{\m\n} R = 8\pi T^{(F)}_{\m\n} \; ,
\lab{einstein-eqs}
\ee
where
\be
T^{(F)}_{\m\n} =
L(F^2)\,G_{\m\n} - 4 L^{\pr}(F^2) F_{\m\k} F_{\n\l} G^{\k\l} \; ,
\lab{T-F}
\ee
and
\be
\pa_\n \(\sqrt{-G} L^{\pr}(F^2)  F_{\k\l} G^{\m\k} G^{\n\l}\)=0 \; ,
\lab{GG-eqs}
\ee
where $L^{\pr} (F^2)$ denotes derivative w.r.t. $F^2$ of the function $L(F^2)$ in 
\rf{gravity+GG}.

In our preceding paper \ct{grav-cornell} we have shown that the gravity-gauge-field
system \rf{gravity+GG} possesses static spherically symmetric solutions
with a radial electric field containing both Coulomb and {\em constant} vacuum
pieces:
\be
F_{0r} = \frac{\vareps_F f}{\sqrt{2}} + \frac{Q}{\sqrt{4\pi}\, r^2} 
\quad ,\quad \vareps_F = \mathrm{sign}(Q) \; ,
\lab{cornell-sol}
\ee
and the space-time metric:  
\br
ds^2 = - A(r) dt^2 + \frac{dr^2}{A(r)} + r^2 \bigl(d\th^2 + \sin^2 \th d\vp^2\bigr)
\; ,
\lab{spherical-static} \\
A(r) = 1 - \sqrt{8\pi}|Q|f - \frac{2m}{r} + \frac{Q^2}{r^2} - \frac{2\pi f^2}{3} r^2 \; ,
\lab{RN-dS+const-electr}
\er
is Reissner-Nordstr{\"o}m-de-Sitter-type with {\em dynamically generated} effective 
cosmological constant $\L_{\mathrm{eff}} = 2\pi f^2$.

Appearance in \rf{RN-dS+const-electr} of a ``leading'' constant term different from 1
resembles the effect on gravity produced by a spherically symmetric ``hedgehog''
configuration of a nonlinear sigma-model scalar field with $SO(3)$ symmetry
\ct{Ed-Rab-hedge} (cf. also \ct{BV-hedge}).

\section{Generalized Levi-Civita-Bertotti-Robinson Space-Times}
\label{gen-BR}
Here we will look for static solutions of Levi-Civita-Bertotti-Robinson type 
\ct{LC-BR-1}--\ct{LC-BR-3} of the system \rf{einstein-eqs}--\rf{GG-eqs}, namely, with 
space-time geometry of the form $\cM_2 \times S^2$ where $\cM_2$ is some two-dimensional
manifold:
\be
ds^2 = - A(\eta) dt^2 + \frac{d\eta^2}{A(\eta)} 
+ r_0^2 \bigl(d\th^2 + \sin^2 \th d\vp^2\bigr) \;\; ,\;\;  
-\infty < \eta <\infty \;\; ,\;\; r_0 = \mathrm{const}
\; ,
\lab{gen-BR-metric}
\ee
and being:
\begin{itemize}
\item
either purely electric type, where the sign factor $\vareps = -1$ in the gauge 
field Lagrangian $L(F^2)$ \rf{gravity+GG}:
\be
F_{\m\n} = 0 \;\; \mathrm{for}\; \m,\n\neq 0,\eta \quad ,\quad
F_{0\eta} = F_{0\eta} (\eta) \; ;
\lab{electr-static}
\ee
\item
or purely magnetic type, where $\vareps = +1$ in \rf{gravity+GG}:
\be
F_{\m\n} = 0 \;\; \mathrm{for}\; \m,\n\neq i,j\equiv \th,\vp \quad ,\quad
\pa_0 F_{ij} = \pa_\vp F_{ij} = 0 \; .
\lab{magnet-static}
\ee
\end{itemize}

In the purely electric case \rf{electr-static} the gauge field equations of
motion become:
\be
\pa_\eta \Bigl( F_{0\eta} - \frac{\vareps_F f}{\sqrt{2}}\Bigr) = 0
\quad ,\quad \vareps_F \equiv \mathrm{sign}(F_{0\eta}) \; ,
\lab{GG-eqs-0}
\ee
yielding a constant vacuum electric field:
\be
F_{0\eta} = c_F = \mathrm{arbitrary ~const} \; .
\lab{const-electr}
\ee
The (mixed) components of energy-momentum tensor \rf{T-F} read:
\be
{T^{(F)}}^0_0 = {T^{(F)}}^\eta_\eta = - \h F^2_{0\eta} \quad ,\quad
T^{(F)}_{ij} = g_{ij}\Bigl(\h F^2_{0\eta} - \frac{f}{\sqrt{2}}|F_{0\eta}|\Bigr)
\; .
\lab{T-F-electr}
\ee
Taking into account \rf{T-F-electr}, the Einstein eqs.\rf{einstein-eqs} for
$(ij)$, where $R_{ij}=\frac{1}{r_0^2} g_{ij}$ because of the $S^2$ factor in
\rf{gen-BR-metric}, yield:
\be
\frac{1}{r_0^2} = 4\pi F^2_{0\eta} \quad , \mathrm{i.e.}\;\; 
r_0 = \frac{1}{2\sqrt{\pi}|c_F|} \; .
\lab{einstein-ij}
\ee
The $(00)$ Einstein eq.\rf{einstein-eqs} using the expression 
$R^0_0 = - \h \pa_\eta^2 A$ (ref.\ct{Ed-Rab-1}; see also \ct{Ed-Rab-2}) becomes:
\be
\pa_\eta^2 A = 8\pi |c_F| \(|c_F| - \sqrt{2}f\) \; .
\lab{einstein-00}
\ee

Therefore, we arrive at the following three distinct types of
Levi-Civita-Bertotti-Robinson solutions for gravity coupled to the
non-Maxwell gauge field system \rf{gravity+GG}:

(i) $AdS_2 \times S^2$ with strong constant vacuum electric field
$|F_{0\eta}| = |c_F|>\sqrt{2}f$, where $AdS_2$ is two-dimensional anti-de Sitter 
space with:
\be
A(\eta) = 4\pi |c_F| \(|c_F| - \sqrt{2}f\)\,\eta^2
\lab{AdS2}
\ee
in the metric \rf{gen-BR-metric}, $\eta$ being the Poincare patch
space-like coordinate.

(ii) $Rind_2 \times S^2$ with constant vacuum electric field 
$|F_{0\eta}| = |c_F|=\sqrt{2}f$, where $Rind_2$ is the flat two-dimensional 
Rindler space with:
\be
A(\eta) = \eta \;\; \mathrm{for}\; 0 < \eta < \infty \quad \mathrm{or} \quad
A(\eta) = - \eta \;\; \mathrm{for}\; -\infty <\eta < 0 
\lab{Rindler2}
\ee
in the metric \rf{gen-BR-metric}.

(iii)  $dS_2 \times S^2$ with weak constant vacuum electric field
$|F_{0\eta}| = |c_F|<\sqrt{2}f$, where $dS_2$ is two-dimensional de Sitter space with:
\be
A(\eta) = 1 - 4\pi |c_F| \(\sqrt{2}f - |c_F|\)\,\eta^2
\lab{dS2}
\ee
in the metric \rf{gen-BR-metric}. For the special value $|c_F| = \frac{f}{\sqrt{2}}$
we recover the Nariai solution \ct{nariai-1,nariai-2} with $A(\eta) = 1 - 2\pi f^2 \eta^2$ 
and equality (up to signs) among energy density, radial and transverse pressures:
$\rho = - p_r = - p_{\perp} = \frac{f^2}{4}$ 
(${T^{(F)}}^\m_\n = \mathrm{diag} \(-\rho,p_r,p_{\perp},p_{\perp}\)$).

In all three cases above the size of the $S^2$ factor is given by 
\rf{einstein-ij}.
Solutions \rf{Rindler2} and \rf{dS2} are new ones and are specifically due to 
the presence of the non-Maxwell square-root term (with $\vareps =-1$) in the gauge 
field Lagrangian \rf{gravity+GG}.

In the purely magnetic case \rf{magnet-static} the gauge field equations of
motion \rf{GG-eqs}:
\be
\pa_\n \Bigl\lb \sin\th \Bigl( 1 + \frac{f}{\sqrt{F^2}}\Bigr) F^{\m\n}\Bigr\rb = 0
\lab{GG-eqs-1}
\ee
yield magnetic monopole solution ~$F_{ij} = B r_0^2\sin\th\,\vareps_{ij}$,
where $B=\mathrm{const}$, irrespective of the presence of the non-Maxwell 
square-root term. However, the latter does contribute to the energy-momentum tensor:
\be
{T^{(F)}}^0_0 = {T^{(F)}}^\eta_\eta = - \h B^2 - f|B| \quad ,\quad
T^{(F)}_{ij} = \h g_{ij} B^2 \; .
\lab{T-F-magnet}
\ee
Taking into account \rf{T-F-magnet}, the Einstein eqs.\rf{einstein-eqs} for $(ij)$
yield (\textsl{cf.} \rf{einstein-ij}):
\be
\frac{1}{r_0^2} = 4\pi\( B^2 + \sqrt{2}f|B|\) \; ,
\lab{einstein-ij-1}
\ee
whereas the mixed-component $(00)$ Einstein eq.\rf{einstein-eqs} gives
$\pa^2_\eta A = 8\pi B^2 $.
Thus in the purely magnetic case we obtain only one solution -- 
$AdS_2 \times S^2$ space-time with magnetic monopole 
where:
\be
A(\eta) = 4\pi B^2 \eta^2
\lab{AdS2-magnet}
\ee
in the metric \rf{gen-BR-metric} and the size of the $S^2$ factor is determined by
\rf{einstein-ij-1}.

\section{Lagrangian Formulation of Lightlike Brane Dynamics}
\label{LL-brane}

In what follows we will consider gravity/gauge-field system
self-consistently interacting with a lightlike $p$-brane (\textsl{LL-brane} for short)
of codimension one ($D=(p+1)+1$). In a series of previous papers
\ct{LL-main-1}--\ct{beograd-2010}
we have proposed manifestly reparametrization invariant world-volume Lagrangian 
formulation in several dynamically equivalent forms of \textsl{LL-branes} coupled to bulk 
gravity $G_{\m\n}$ and bulk gauge fields, in particular, electromagnetic field $A_\m$. 
Here we will use our Polyakov-type formulation given by the world-volume action:
\br
S_{\rm LL}\lb q\rb  = - \h \int d^{p+1}\s\, T b_0^{\frac{p-1}{2}}\sqrt{-\g}
\llb \g^{ab} {\bar g}_{ab} - b_0 (p-1)\rrb \; ,
\lab{LL-action+EM} \\
{\bar g}_{ab} \equiv \pa_a X^\m G_{\m\n} \pa_b X^\n 
- \frac{1}{T^2} (\pa_a u + q\cA_a)(\pa_b u  + q\cA_b) 
\quad , \quad \cA_a \equiv \pa_a X^\m A_\m \; .
\lab{ind-metric-ext-A}
\er
Here and below the following notations are used:
\begin{itemize}
\item
$\g_{ab}$ is the {\em intrinsic} Riemannian metric on the world-volume with
$\g = \det \Vert\g_{ab}\Vert$;
$g_{ab}$ is the {\em induced} metric on the world-volume:
\be
g_{ab} \equiv \pa_a X^{\m} G_{\m\n}(X) \pa_b X^{\n} \; ,
\lab{ind-metric}
\ee
which becomes {\em singular} on-shell (manifestation of the lightlike nature), 
\textsl{cf.} Eq.\rf{on-shell-singular-A} below); 
$b_0$ is a positive constant measuring the world-volume ``cosmological constant''.
\item
$X^\m (\s)$ are the $p$-brane embedding coordinates in the bulk
$D$-dimensional space-time with Riemannian metric
$G_{\m\n}(x)$ ($\m,\n = 0,1,\ldots ,D-1$); 
$(\s)\equiv \(\s^0 \equiv \t,\s^i\)$ with $i=1,\ldots ,p$;
$\pa_a \equiv \partder{}{\s^a}$.
\item
$u$ is auxiliary world-volume scalar field defining the lightlike direction
of the induced metric 
(see Eq.\rf{on-shell-singular-A} below) and it is a non-propagating degree of freedom
( ref.\ct{beograd-2010}).
\item
$T$ is {\em dynamical (variable)} brane tension (also a non-propagating
degree of freedom). 
\item
Coupling parameter $q$ is the surface charge density of the \textsl{LL-brane}.
\end{itemize}

The corresponding equations of motion w.r.t. $X^\m$, $u$, $\g_{ab}$ and $T$ read accordingly
(using short-hand notation \rf{ind-metric-ext-A}):
\br
\pa_a \( T \sqrt{|{\bar g}|} {\bar g}^{ab}\pa_b X^\m\)
+ T \sqrt{|{\bar g}|} {\bar g}^{ab} \pa_a X^\l \pa_b X^\n \G^\m_{\l\n}
\nonu \\
+ \frac{q}{T} \sqrt{|{\bar g}|} {\bar g}^{ab}
\pa_a X^\n (\pa_b u  + q\cA_b) \cF_{\l\n}G^{\m\l} = 0 \; ,
\lab{X-eqs-NG-A} \\
\pa_a \(\frac{1}{T} \sqrt{|{\bar g}|} {\bar g}^{ab}(\pa_b u  + q\cA_b)\) = 0
\quad ,\quad  
\g_{ab} = \frac{1}{b_0} {\bar g}_{ab}  \; ,
\lab{u-gamma-eqs-NG-A} \\
T^2 + \eps {\bar g}^{ab}(\pa_a u  + q\cA_a)(\pa_b u  + q\cA_b) = 0 \; .
\lab{T-eq-NG-A}
\er
Here ${\bar g} = \det\Vert {\bar g}_{ab} \Vert$ and $\G^\m_{\l\n}$ denotes the Christoffel 
connection for the bulk metric $G_{\m\n}$.

The on-shell singularity of the induced metric $g_{ab}$ \rf{ind-metric}, \textsl{i.e.}, 
the lightlike property, directly follows Eq.\rf{T-eq-NG-A} and the definition of
${\bar g}_{ab}$ \rf{ind-metric-ext-A}:
\be
g_{ab} \({\bar g}^{bc}(\pa_c u  + q\cA_c)\) = 0 \; .
\lab{on-shell-singular-A}
\ee

Explicit world-volume reparametrization invariance of the \textsl{LL-brane} action
\rf{LL-action+EM} allows to introduce the standard synchronous gauge-fixing conditions
for the intrinsic world-volume metric  
$\g_{00} = -1\; ,\; \g_{0i} = 0 \; (i=1,\ldots,p)$.
which reduces Eqs.\rf{u-gamma-eqs-NG-A}--\rf{T-eq-NG-A} to the following relations:
\br
\frac{(\pa_0 u + q\cA_0)^2}{T^2} = b_0 + g_{00} \quad ,\quad 
\pa_i u + q\cA_i= (\pa_0 u + q\cA_0) g_{0i} \( b_0 + g_{00}\)^{-1} \; ,
\nonu \\
g_{00} = g^{ij} g_{0i} g_{0j} \;\; ,\;\;
\pa_0 \(\sqrt{g^{(p)}}\) + \pa_i \(\sqrt{g^{(p)}}g^{ij} g_{0j}\) = 0 \;\; ,\;\; 
g^{(p)} \equiv \det\Vert g_{ij}\Vert \; ,
\lab{g-rel}
\er
(recall that $g_{00},g_{0i},g_{ij}$ are the components of the induced metric 
\rf{ind-metric}; $g^{ij}$ is the inverse matrix of $g_{ij}$).
Then, as shown in refs.\ct{LL-main-1}--\ct{beograd-2010}, 
consistency of \textsl{LL-brane} dynamics in static
``spherically-symmetric''-type backgrounds (in what follows we will use 
Eddington-Finkelstein coordinates, $dt=dv-\frac{d\eta}{A(\eta)}$):
\be
ds^2 = - A(\eta) dv^2 + 2dv d\eta + C(\eta) h_{ij}(\th) d\th^i d\th^j \quad ,\quad
F_{v\eta} = F_{v\eta} (\eta)\; ,\; \mathrm{rest}=0
\lab{static-spherical-EF}
\ee
with the standard embedding ansatz:
\be
X^0\equiv v = \t \quad, \quad X^1\equiv \eta = \eta (\t) \quad, \quad 
X^i\equiv \th^i = \s^i \;\; (i=1,\ldots ,p) \; .
\lab{X-embed}
\ee
requires the corresponding background \rf{static-spherical-EF} to possess a horizon 
at some $\eta\!=\!\eta_0$, which is automatically occupied by the \textsl{LL-brane},
\textsl{i.e.}:
\be
\eta (\t) = \eta_0 \quad ,\quad A(\eta_0) = 0 \; .
\lab{straddling}
\ee
This property is called ``horizon straddling'' according to the terminology
of Ref.\ct{Barrabes-Israel}. Similar ``horizon straddling'' has been found also for 
\textsl{LL-branes} moving in rotating axially symmetric (Kerr or Kerr-Newman) and 
rotating cylindrically symmetric black hole backgrounds 
\ct{Kerr-rot-WH-1,Kerr-rot-WH-2}. 

\section{Self-Consistent Wormhole-Like Solutions with LL-Branes}
\label{self-consist}

Let us now consider a bulk gravity/gauge-field system  in $D\!=\!4$
\rf{gravity+GG} self-consistently interacting with a $p=2$ \textsl{LL-brane}:
\be
S = \int d^4 x \sqrt{-G} \Bigl\lb \frac{R(G)}{16\pi} 
 - \frac{1}{4} F^2 - \frac{f}{2} \sqrt{- F^2}\Bigr\rb + S_{\rm LL}\lb q\rb \; ,
\lab{gravity+GG+LL}
\ee
where $S_{\rm LL}\lb q\rb$ is the \textsl{LL-brane} world-volume action 
\rf{LL-action+EM} (with $p=2$). It is now the \textsl{LL-brane} which will be the material and
charge source for gravity and (nonlinear) electromagnetism.

The equations of motion resulting from \rf{gravity+GG+LL} read:
\br
R_{\m\n} - \h G_{\m\n} R = 
8\pi \llb T^{(F)}_{\m\n} + T^{(\mathrm{brane})}_{\m\n}\rrb \; ,
\lab{einstein+LL-eqs}\\
\pa_\n \Bigl\lb\sqrt{-g} \Bigl( 1 - \frac{f}{\sqrt{-F^2}}\Bigr) 
F_{\k\l} G^{\m\k} G^{\n\l}\Bigr\rb + j_{(\mathrm{brane})}^\m = 0 \; ,
\lab{GG+LL-eqs}
\er
together with the \textsl{LL-brane} equations \rf{X-eqs-NG-A}--\rf{T-eq-NG-A}.
$T^{(F)}_{\m\n}$ is the same as in \rf{T-F}.
The energy-momentum tensor and the charge current density of the \textsl{LL-brane} are 
straightforwardly derived from the underlying world-volume action \rf{LL-action+EM}:
\br
T_{(\mathrm{brane})}^{\m\n} = 
- \int\!\! d^3\s\,\frac{\d^{(4)}\Bigl(x-X(\s)\Bigr)}{\sqrt{-G}}
\, T\,\sqrt{|{\bar g}|} {\bar g}^{ab} \pa_a X^\m \pa_b X^\n \; ,
\lab{T-brane-A} \\
j_{(\mathrm{brane})}^\m = - q \int\!\! d^3\s\,\d^{(4)}\Bigl(x-X(\s)\Bigr)
\sqrt{|{\bar g}|} {\bar g}^{ab}\pa_a X^\m \(\pa_b u + q\cA_b\)T^{-1} \; .
\lab{j-brane-A}
\er

Looking for solutions of static ``spherically-symmetric''-type \rf{static-spherical-EF}
for the coupled gravity-gauge-field-\textsl{LL-brane} system \rf{gravity+GG+LL}
amounts to the following simple steps:

(i) Choose ``vacuum'' static ``spherically-symmetric''-type solutions
\rf{static-spherical-EF} of \rf{einstein+LL-eqs}--\rf{GG+LL-eqs} (\textsl{i.e.}, 
without the delta-function terms due to the \textsl{LL-branes}) in each region 
$-\infty < \eta < \eta_0$ and $\eta_0 <\eta < \infty$ with a common horizon
at $\eta=\eta_0$;

(ii) The \textsl{LL-brane} automatically locates itself on the horizon
according to ``horizon straddling'' property \rf{straddling};

(iii) Match the discontinuities of the derivatives of the metric and
the gauge field strength \rf{static-spherical-EF} across the horizon at $\eta = \eta_0$
using the explicit expressions for the \textsl{LL-brane} stress-energy tensor 
charge current density \rf{T-brane-A}--\rf{j-brane-A}.

Using \rf{g-rel}--\rf{X-embed} we find for the \textsl{LL-brane} energy-momentum 
tensor and charge current density:
\be
T_{(\mathrm{brane})}^{\m\n} = S^{\m\n}\,\d (\eta-\eta_0) \quad ,\quad
j_{(\mathrm{brane})}^\m = \d^\m_0 q\sqrt{\det\Vert G_{ij}\Vert}\, \d (\eta-\eta_0) \; ,
\lab{T-j-0}
\ee
where $G_{ij} = C(\eta) h_{ij}(\th)$ (\textsl{cf.} \rf{static-spherical-EF}).
The non-zero components of the surface energy-momentum tensor $S_{\m\n}$ (with lower indices)
and its trace are:
\be
S_{\eta\eta} = \frac{T}{b_0^{1/2}} \quad ,\quad 
S_{ij} = - T b_0^{1/2} G_{ij} \quad ,\quad 
S^\l_\l = - 2Tb_0^{1/2} \; .
\lab{S-comp}
\ee
Taking into account \rf{T-j-0}--\rf{S-comp} together with
\rf{static-spherical-EF}--\rf{straddling}, the matching relations at the
horizon $\eta =\eta_0$ become \ct{BR-kink}--\ct{beograd-2010} 
(for a systematic introduction to the formalism of matching different bulk
space-time geometries on codimension-one hypersurfaces (``thin shells'') see
the textbook \ct{poisson-kit}):

(A) Matching relations from Einstein eqs.\rf{einstein+LL-eqs}:
\be
\llb \pa_\eta A \rrb_{\eta_0} = - 16\pi T \sqrt{b_0} 
\quad,\quad 
\llb \pa_\eta \ln C \rrb_{\eta_0} = - \frac{8\pi}{\sqrt{b_0}} T
\lab{eqsys-1-2}
\ee
with notation $\bigl\lb Y \bigr\rb_{\eta_0} \equiv 
Y\bv_{\eta \to \eta_0 +0} - Y\bv_{\eta \to \eta_0 -0}$ for any quantity $Y$.

(B) Matching relation from nonlinear gauge field eqs.\rf{GG+LL-eqs}:
\be
\llb F_{v\eta} \rrb_{\eta_0} = q
\lab{eqsys-4}
\ee

(C) $X^0$-equation of motion of the \textsl{LL-brane} (the only non-trivial
contribution of second-order \textsl{LL-brane} eqs.\rf{X-eqs-NG-A} in the
case of embedding \rf{X-embed}):
\be
\frac{T}{2} \( \llangle \pa_\eta A \rrangle_{\eta_0} 
+ 2 b_0 \llangle \pa_\eta \ln C \rrangle_{\eta_0} \)
-  \sqrt{b_0} q \llangle F_{v\eta}\rrangle_{\eta_0} = 0
\lab{eqsys-3}
\ee
with notation $\llangle Y \rrangle_{\eta_0} \equiv 
\h \( Y\bv_{\eta \to \eta_0 +0} + Y\bv_{\eta \to \eta_0 -0}\)$. 

We are looking for wormhole-type solutions to \rf{gravity+GG+LL} with
the charged \textsl{LL-brane} at the wormhole ``throat'' connecting a non-compact 
``universe'' with Reissner-Nordstr{\"o}m-de-Sitter-type geometry 
\rf{cornell-sol}--\rf{RN-dS+const-electr} (where the cosmological constant is
{\em dynamically} generated) to a compactified (``tube-like'') 
``universe'' of (generalized) Levi-Civita-Bertotti-Robinson type 
\rf{gen-BR-metric}--\rf{electr-static}. These wormholes possess
the novel property of {\em hiding} electric charge from external observer in the
non-compact ``universe'', \textsl{i.e.}, the whole electric flux produced by
the charged \textsl{LL-brane} at the wormhole ``throat'' is pushed into the 
``tube-like'' ``universe''.

The first wormhole-type solution of the above kind we find is given by:

(a) ``left universe'' of Levi-Civita-Bertotti-Robinson (``tube-like'') type 
with geometry $Rind_2 \times S^2$ \rf{Rindler2}:
\be
A(\eta) = - \eta \quad ,\quad C(\eta) = r_0^2 \quad ,\quad
|F_{v\eta}| = \sqrt{2}f \quad \mathrm{for}\;\; \eta< 0 \; ;
\lab{BR-Rind-left}
\ee

(b) non-compact ``right universe'' comprising the exterior region of 
Reissner-Nordstr{\"o}m-de-Sitter-type black hole beyond the middle (Schwarzschild-type) 
horizon $r_0$ (\textsl{cf.} \rf{cornell-sol}--\rf{RN-dS+const-electr}):
\br
A(\eta) = 1 - \sqrt{8\pi}|Q|f  - \frac{2m}{r_0 + \eta} + \frac{Q^2}{(r_0 + \eta)^2} 
- \frac{2\pi f^2}{3} (r_0 + \eta)^2 \; , 
\nonu \\
A(0) = 0 \;,\;\pa_\eta A(0) > 0 \; , \phantom{aaaaaaaaaaaaa}
\nonu \\
C(\eta) = (r_0 + \eta)^2 \quad ,\quad
F_{v\eta} = \frac{\vareps_F f}{\sqrt{2}} + \frac{Q}{\sqrt{4\pi}\, (r_0 + \eta)^2} 
\quad \mathrm{for}\;\; \eta> 0 \; .
\lab{RNdS-right}
\er

Substituting \rf{BR-Rind-left}--\rf{RNdS-right} into the set of matching
relations \rf{eqsys-1-2}--\rf{eqsys-3} determines all parameters of the wormhole 
$\( r_0, m, Q, b_0, q\)$ in terms of the coupling constant $f$ in front of
the square-root Maxwell term in \rf{gravity+GG+LL}:
\br
Q=0 \;\; ,\;\; |q| = \frac{f}{\sqrt{2}} \;\;,\;\; 
\mathrm{sign}(q) = - \mathrm{sign}(F_{v\eta}) \; ,
\lab{param-1} \\
r_0^2 = \frac{1}{8\pi f^2} \;\; ,\;\; m = \frac{11}{48\sqrt{2\pi}\,f} \;\; ,\;\;
b_0 = \frac{1}{8\sqrt{2\pi}\,f} + \frac{3}{16} \; .
\lab{param-2}
\er

The second wormhole-type solution of the aforementioned kind reads:

(c) ``left universe'' of Levi-Civita-Bertotti-Robinson (``tube-like'') type 
with geometry $AdS_2 \times S^2$ \rf{AdS2}:
\br
A(\eta) = 4\pi |c_F| \(|c_F| - \sqrt{2}f\)\,\eta^2 
\quad ,\quad C(\eta) = r_0^2 \; ,
\nonu\\
|F_{v\eta}| = |c_F| > \sqrt{2}f \quad \mathrm{for}\;\; \eta< 0 \; ;
\lab{BR-AdS2-left}
\er

(d) non-compact Reissner-Nordstr{\"o}m-de-Sitter-type ``right universe'' of the same 
kind as \rf{RNdS-right}.

Substituting again \rf{BR-AdS2-left}, \rf{RNdS-right} into the matching
relations \rf{eqsys-1-2}--\rf{eqsys-3} we find for the wormhole parameters:
\br
Q=0 \;\; ,\;\; |c_F| = |q| + \frac{f}{\sqrt{2}} \;\;,\;\; 
\mathrm{sign}(q) = - \mathrm{sign}(F_{v\eta}) \equiv - \mathrm{sign}(c_F) \; ,
\lab{param-3}\\
r_0^2 = \frac{1}{4\pi c_F^2} \;\; ,\;\; 
m = \frac{1}{2\sqrt{\pi}\,f}\Bigl( 1 - \frac{f^2}{6 c_F^2}\Bigr) \;\; ,\;\;
b_0 = \frac{|q|\(|q|+\sqrt{2}f\)}{4 c_F^2} \; .
\lab{param-4}
\er

The important observation here is that $Q=0$ in both wormhole solutions 
(a)-(b) (Eqs.\rf{BR-Rind-left}--\rf{RNdS-right}, \rf{param-1}--\rf{param-2}) 
and (c)-(d) (Eqs.\rf{BR-AdS2-left}, \rf{RNdS-right}, \rf{param-3}--\rf{param-4}). 
Therefore, the ``right universe'' in both cases turns out 
to be the exterior region of the electrically neutral {\em Schwarzschild-de-Sitter}
black hole beyond the Schwarzschild horizon which carries a vacuum constant radial
electric field $|F_{v\eta}| = \frac{f}{\sqrt{2}}$. On the other hand, according to
\rf{RNdS-right},\rf{param-1} and \rf{RNdS-right},\rf{param-3} the whole flux 
produced by the \textsl{LL-brane} charge $q$ ($|F_{v\eta}|=\frac{f}{\sqrt{2}}+|q|$)
flows only into the compactified ``left universe'' of 
Levi-Civita-Bertotti-Robinson type ($Rind_2 \times S^2$ \rf{Rindler2} or 
$AdS_2 \times S^2$ \rf{AdS2}).

The geometry of the above constructed charge-``hiding'' wormhole solutions
is illustrated in Figure 1.

\begin{figure}
\begin{center}
\includegraphics{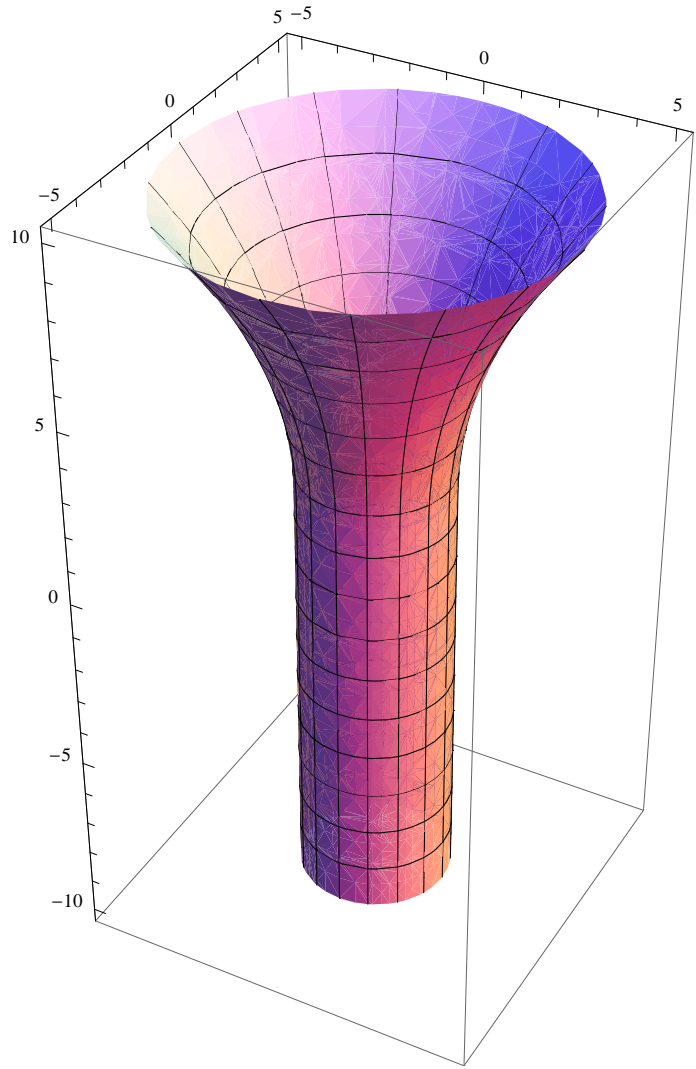}
\caption{Shape of $t=const$ and $\th=\frac{\pi}{2}$ slice of
charge-``hiding'' wormhole geometry. The whole electric flux is expelled into the
lower (infinitely long) cylindric tube.}
\end{center}
\end{figure}

\section{Conclusions}
\label{conclude}
We have seen that a charged wormhole ``throat'' realized by a charged
lightlike brane, when joining a compactified space-time with a non-compact
space-time region, expels all of the electric flux it produces into the
compactified (``tube-like'') region when the gauge field dynamics is driven
by an additional ``square-root'' Maxwell term known to produce QCD-like
confining potential in flat space-time. Indeed, this effect can be
understood from the point of view of an observer in the non-compact
``universe'' as an alternative way of achieving charge confinement in a
fashion similar to the MIT bag model \ct{MIT-bag}, where the role of the inside bag 
region is being played by the compactified Levi-Civita-Bertotti-Robinson
space.

In an accompanying paper \ct{hide-confine} we show that the above ``charge-hiding'' 
solution can be further generalized to a truly
{\em charge-confining} wormhole solution when we couple the bulk
gravity/nonlinear-gauge-field system \rf{gravity+GG} self-consistently to {\em two} separate
codimension-one charged {\em lightlike} branes with equal in magnitude but opposite 
charges. The latter system possesses a ``two-throat'' wormhole solution, where the 
``left-most'' and the ``right-most'' ``universes''
are two identical copies of the exterior region of the neutral Schwarzschild-de-Sitter 
black hole beyond the Schwarzschild horizon,
whereas the ``middle'' ``universe'' is of generalized Levi-Civita-Bertotti-Robinson
``tube-like'' form with geometry $dS_2 \times S^2$ \rf{dS2}.
It comprises the finite-extent intermediate region of $dS_2$ between its two horizons. Both
``throats'' are occupied by the two oppositely charged lightlike branes and the whole
electric flux produced by the latter is confined entirely within the middle
finite-extent ``tube-like'' ``universe'' 
-- a property qualitatively resembling the
quark confinement phenomenon in quantum chromodynamics.

\section*{Acknowledgments}
E.N. and S.P. are supported by Bulgarian NSF grant \textsl{DO 02-257}.
Also, all of us acknowledge support of our collaboration through the exchange
agreement between the Ben-Gurion University 
and the Bulgarian Academy of Sciences. 
We are grateful to Stoycho Yazadjiev for constructive discussions. We also
thank Doug Singleton for correspondence.

\end{document}